\documentclass[11pt]{article}
\begin{document}
\title{Correlation Functions and AdS/LCFT Correspondence}
\author{S. Moghimi-Araghi \footnote{e-mail: samanimi@mehr.sharif.edu}
, S. Rouhani \footnote{e-mail: rouhani@karun.ipm.ac.ir} and M.
Saadat \footnote{e-mail: msaadat@sina.sharif.edu}\\ \\
Department of Physics, Sharif University of Technology,\\ Tehran,
P.O.Box: 11365-9161, Iran} \maketitle

\begin{abstract}
Correlation functions of Logarithmic conformal field theory is
investigated using the ADS/CFT correspondence and a novel method
based on nilpotent weights and 'super fields'. Adding an specific
form of interaction, we introduce a perturbative method to
calculate the correlation
functions. \vspace{5mm}%
\end{abstract}
\section{Introduction}
In this paper, we will examine calculation of correlation
functions in a logarithmic conformal field theory, which has been
derived using the AdS/CFT correspondence.  The correspondence
which is by now very popular in the string theory literature,
maintains that a $d$ dimensional conformal theory may arise on the
boundary of a $d+1$ dimensional theory defined on an Anti de
Sitter back ground \cite{maldacena}. Since Logarithmic Conformal
Field Theories (LCFTs) are singular forms of CFTs, one expects
that they arise on the boundary too, but perhaps as boundary
theories of rather unusual bulk theories \cite{Khorrami,KoganAds}.
The goal of the present discussion is to examine the way one loop
correlation functions within LCFT may be calculated, if the bulk
theory is given.  A method based on Feynman graphs which connect
to the boundary, has been suggested by Witten \cite{Witten}, which
gives the road map by which correlation functions for the boundary
CFT can be calculated. The same method works for LCFTs as well.
Except that the number of correlation function to calculate for an
LCFT is very high. In the simplest case, where one has a
logarithmic pair, there are sixteen interconnected correlation
functions when considering four point functions. Recently we
proposed a method for dealing with LCFTs based on nilpotent
variables \cite{MRS}. This method offers a unified manner by which
one can calculated the correlation functions of LCFTs
\cite{MRSAlgeb}, and easily generalizes to the case of AdS/LCFT
correspondence \cite{MRSAdS}. However its application to the
graphs with loops was riddled with some difficulties
\cite{JabRouh}. Here we present a method by which these problems
are ironed out and we are able to discuss the properties of loop
calculations.

\section{AdS/LCFT Correspondence}
 To state the AdS/LCFT correspondence, we briefly recall the
conjecture in the frame work of ordinary CFT. In this conjecture,
one relates two different theories: one of them lives in a $d+1$
dimensional AdS space, and the other somehow lives on the boundary
of this AdS space and has conformal invariance. Let us be more
precise. Consider the action $S[\Phi]$ defined on AdS$_{d+1}$ and
calculate the partition function of the AdS theory subjected to
the constraint that the value of $\Phi$ on the boundary be
$\Phi_{b}$
\begin{equation}
Z_{AdS}[\Phi_{b}]=\int_{\Phi_{b}} D\Phi\exp(-S[\Phi]).
\end{equation}
The correspondence states that the partition function of AdS
theory is the generating functional of the boundary conformal
field theory:
\begin{equation}\label{eq:a2}
Z_{AdS}[\Phi_{b}]=\left\langle \exp\left(\int_{\partial AdS} dx
\hat{O}\Phi_b\right)\right\rangle.
\end{equation}
The function $\Phi_b$ is considered as a current which couples to
the scalar conformal operator $\hat{O}$ via  a coupling
$\int_{\partial AdS}dx \hat{O}\Phi_b$. This is an elegant and
useful result, since it gives a practical way for calculation of
correlation functions of conformal field theory. To begin, one
should first choose a proper action in the bulk. Different actions
have been studied and the conformal correlators have been found in
these cases. Some examples are interacting massive scalar field
theory \cite{Muck} or interacting scalar spinor field theory
\cite{hening}.

On the other hand, it is also interesting to find actions on AdS
which induce logarithmic conformal field theory (LCFT) on its
boundary.

The bulk action which give rise to logarithmic operators on the
boundary were first described in \cite{Khorrami,KoganAds}. Then it
was accustomed to the new method for investigating LCFTs, i.e. via
nilpotent and grassmann variables \cite{MRSAdS}. In this new
method a superfield is defined using a grassmannian variable
$\eta$ and different components of a logarithmic pair and
fermionic fields
\begin{equation}\label{eq:a4}
\hat{O}(\vec{x},\eta,\bar{\eta})=\hat{A}(\vec{x})+\hat{\bar{\zeta}}
(\vec{x})\eta+\bar{\eta}\hat{\zeta}(\vec{x})+\bar{\eta}\eta\hat{B}(\vec{x})
\end{equation}
where $\hat{\zeta}(\vec{x})$ and $\hat{\bar{\zeta}}(\vec{x})$ are
fermionic fields with the same conformal dimension as
$\hat{A}(\vec{x})$, and $\hat{B}(\vec{x})$ is the logarithmic
partner of $\hat{A}(\vec{x})$. Also $\bar{\eta}\eta$ acts as the
nilpotent variable. Now it is easy to see that
$\hat{O}(\vec{x},\eta,\bar{\eta})$ has the following
transformation law
\begin{equation}\label{eq:a5}
\hat{O}(\lambda\vec{x},\eta,\bar{\eta})=\lambda^{-(\Delta+\bar{\eta}\eta)}
\hat{O}(\vec{x},\eta,\bar{\eta
})
\end{equation}
If $ \hat{O}(\vec{x},\eta,\bar{\eta})$ were the logarithmic
operator on the boundary of AdS, the corresponding field
$\Phi(x,\eta,\bar{\eta})$ in AdS space can be extended as
\begin{equation}
\Phi(x,\eta,\bar{\eta})=C(x)+\bar{\eta}\alpha(x)+\bar{\alpha}(x)\eta+
\bar{\eta}\eta D(x)
\end{equation}
where $x$ is ($d+1$) dimensional coordinate with
$x^{0},x^{1},...,x^{d}$ components and $ x=(\vec{x},x^{d})$ and
$x^{d}=0$ corresponds to the
  boundary.
In reference \cite{MRSAdS} we introduced a free action with BRST
symmetry. Now we write it as the following form
\begin{eqnarray}\label{FreeAction}
S^f[\Phi]=-\frac{1}{2}\int_{\Omega}d^{d+1}x\sqrt{|g(x)|}\int
d\bar{\eta}d\eta\times\:\hspace{5cm}\nonumber\\
\left[\vec{\nabla}\Phi(x,\eta,\bar{\eta}).
\vec{\nabla}\Phi(x,-\eta,-\bar{\eta})+m^{2}(\eta,\bar{\eta})
\Phi(x,\eta,\bar{\eta})\Phi(x,-\eta,-\bar{\eta})\right]
\end{eqnarray}
where $m^{2}(\eta,\bar{\eta})=m_{1}^{2}+m_{2}^{2}\bar{\eta}\eta$
and $m_{1}^{2}$, $m_{2}^{2}$ are defined to be $\Delta(\Delta-d)$
and $(2\Delta-d)$ respectively and where $|g(x)|$ is the
determinant of the metric on AdS.

Within this framework, the AdS/LCFT correspondence is written in
the following form
\begin{equation}\label{correspondence}
\left\langle\exp\int d\bar{\eta}d\eta\int_{\partial
\Omega}d^{d}\vec{x}
\hat{O}(\vec{x},\eta,\bar{\eta})\Phi_{b}(\vec{x},\eta,\bar{\eta})\right\rangle
= \exp\left(-S_{cl}[\Phi_{b}(\vec{x},\eta,\bar{\eta})]\right),
\end{equation}
where $\Phi_{b}(\vec{x},\eta,\bar{\eta})$ is the value of
$\Phi(x,\eta,\bar{\eta})$ on the boundary. Then the two point
correlation functions were calculated \cite{MRSAdS}, though the
results were not in closed form and one had to expand
$\Phi(x,\eta)$ in terms of $C(x)$, $\alpha(x)$, $\bar{\alpha}(x)$
and $D(x)$ to obtain correlation function.

To have non trivial 4 or higher point functions, one should add
some kind of interaction to the action (\ref{FreeAction}). In
\cite{JabRouh} various n-point correlations of superfield
components were calculated at tree level by adding an interaction
term to the free theory. Again, the results the correlation
functions were not found in closed form. Also they had chosen a
special form for interaction, more precisely, they have taken the
interaction term to be proportional to $\prod_1^n \phi(x,\eta_i)$,
but to keep the action BRST invariance, the sum  of $\eta_i$
should be zero. The choice made in \cite{JabRouh} is to take half
of $\eta_i$'s equal to $\eta$ and the other half equal to
$-\eta$.\footnote{Of course this can be done only when $n$ is
even, if $n$ is odd, another choice should be made.} In this paper
we will not restrict ourselves to a specific choice and will
consider more general cases. It turns out that with the new
action, nearly all the correlators will have correction, while in
\cite{JabRouh} only few term had correction. It is also important
to see what happens when one goes beyond tree level. In ordinary
CFT's, Witten diagrams have been introduced to calculate n-point
functions, perturbatively. We will generalize this idea to LCFT's
and will find proper Feynman rules to describe the diagrams.

In this paper we obtain n-point correlation functions of
superfields in closed form instead of its components. This method
leads us to generalization of Witten diagrams which deal with
ordinary fields to the diagrams dealing with superfields. Also by
this generalization we can evaluate loop corrections.
\section{Interaction action and the generalized Witten diagrams}
In order to find non trivial n-point functions at tree level and
loop corrections, one should consider interaction terms in
$\Phi(x,\eta,\bar{\eta})$ in addition to free theory. Furthermore
we wish the action to have BRST symmetry so that the correlation
functions remain invariant under BRST transformations. We consider
the interaction term as
\begin{eqnarray}\label{interactionaction}
S^I[\Phi]&=&\int_{\Omega}d^{d+1}x\sqrt{|g(x)|}\int
d\bar{\eta}_{1}d\eta_{1}\cdots
d\bar{\eta}_{n}d\eta_{n}\delta(\eta_{1}+\cdots+\eta_{n})\delta(\bar{\eta}_{1}+\cdots+\bar{\eta}_{n})\nonumber\\
&&\left[\frac{1}{n!}\lambda(\eta_{1},\cdots,\eta_{n},\bar{\eta}_{1},\cdots,\bar{\eta}_{n})
\Phi(x,\eta_{1},\bar{\eta}_{1})\cdots\Phi(x,\eta_{n},\bar{\eta}_{n})\right].
\end{eqnarray}

In the language of superfield, the infinitesimal BRST
transformation is of the form
\begin{equation}
\delta\Phi(x,\eta,\bar{\eta})=(\bar{\epsilon}\eta+\bar{\eta}\epsilon)\Phi(x,\eta,\bar{\eta}),
\end{equation}
where $\bar{\epsilon}$ and $\epsilon$ are infinitesimal
anti-commuting parameters. Presence of Dirac delta functions in
$S^{I}[\Phi]$ guarantees BRST symmetry of these actions, because
\begin{equation}
\delta\left[\Phi(x,\eta_{1},\bar{\eta}_{1})\cdots\Phi(x,\eta_{n},\bar{\eta}_{n})\right]=
\left[\Phi(x,\eta_{1},\bar{\eta}_{1})\cdots\Phi(x,\eta_{n},\bar{\eta}_{n})\right]
\left[\bar{\epsilon}\sum_{i=1}^{n}\eta_{i}+\sum_{i=1}^{n}\bar{\eta}_{i}\epsilon\right].
\end{equation}

With this interaction term, we will move on to establish suitable
Fenyman rules for generalized Witten diagrams. To have all the
rules, we should study three terms: first, the bulk-bulk Green
function, second, the surface-bulk Green function and third the
vertex related to the interaction term (\ref{interactionaction}).

As we will see, it is better to rewrite the free action (equation
(\ref{FreeAction})) in the following form, which is a more
suitable form to derive Feynman rules,
\begin{eqnarray}
S^f[\Phi]=-\frac{1}{2}\int_{\Omega}d^{d+1}x\sqrt{|g(x)|}\int
d\bar{\eta}_{1}d\eta_{1}\:d\bar{\eta}_{2}d\eta_{2}\delta(\eta_{1}+\eta_{2})\delta(\bar{\eta}_{1}+\bar{\eta}_{2})\nonumber\\
\left[\vec{\nabla}\Phi(x,\eta_{1},\bar{\eta}_{1}).
\vec{\nabla}\Phi(x,\eta_{2},\bar{\eta}_{2})+m^{2}(\eta_{1},\eta_{2},\bar{\eta}_{1},\bar{\eta}_{2})
\Phi(x,\eta_{1},\bar{\eta}_{1})\Phi(x,\eta_{2},\bar{\eta}_{2})\right].
\end{eqnarray}

From now on we drop $\bar{\eta}$ dependence of all entities except
integrals measures for simplicity. Considering total action
$S=S^{f}[\Phi]+S^{I}[\Phi]$, $\frac{\delta S}{\delta
\Phi(x,\eta_{1})}$=0 leads the equation of motion
\begin{eqnarray}\label{eq:eqationmotion}
&&\left(\nabla^{2}-m^{2}(\eta_{1})\right)\Phi(x,\eta_{1})=-\int
d\bar{\eta}_{2}d\eta_{2}\cdots d\bar{\eta}_{n}d\eta_{n}
\delta(\eta_{1}+\cdots+\eta_{n})\nonumber\\
&&\hspace{4cm}\frac{1}{(n-1)!}\lambda(\eta_{1},\cdots,\eta_{n})\Phi(x,\eta_{2})\cdots\Phi(x,\eta_{n}).
\end{eqnarray}
Introducing Dirichlet Green function for this system as
\begin{eqnarray}\label{greenfunction}
\left(\nabla^{2}-m^{2}(\eta_{1})\right){\mathcal
G}(x,y,\eta_{1},\eta_{2})=
\frac{\delta(x-y)\delta(\eta_{1}+\eta_{2})}{\sqrt{|g(x)|}},
\end{eqnarray}
together with the boundary condition
\begin{eqnarray}
\int d\bar{\eta}_{2}d\eta_{2} {\mathcal G}
(x,y,\eta_{1},\eta_{2})|_{x\in\partial\Omega}=0,
\end{eqnarray}
and applying Green's theorem we obtain
\begin{eqnarray}\label{motionsul}
\Phi(x,\eta_{1})&=&\int d\bar{\eta}_{2}d\eta_{2}\int
d^{d+1}y\sqrt{|g(y)|}\:{\mathcal G}(x,y,\eta_{1},\eta_{2})
\left(\nabla^{2}-m^{2}(\eta_{2})\right)\Phi(y,\eta_{2})\nonumber\\
&+&\int
d\bar{\eta}_{2}d\eta_{2}\int_{\partial\Omega}d^{d}y\sqrt{|h(y)|}
\:\Phi(y,\eta_{2})n^{\mu}\frac{\partial}{\partial
y^{\mu}}{\mathcal G}(x,y,\eta_{1},\eta_{2})
\end{eqnarray}
where $|h(x)|$ is the determinant of the induced metric on
$\partial\Omega$ and $n^{\mu}$ the unit vector normal to
$\partial\Omega$ and pointing outwards. Note that because of the
new form of free action, the Green function has both dependence on
$\eta_1$ and $\eta_2$. This Green function is just the one derived
in \cite{MRSAdS} multiplied by
$\delta(\eta_1+\eta_2)=\eta_1+\eta_2$. That is
\begin{equation}
{\mathcal G}(x,y,\eta_1,\eta_2)=
G(x,y,\eta_1)\delta(\eta_1+\eta_2),
\end{equation}
where $G$ is introduced in \cite{MRSAdS}. Note that as the
dependence of this Green function on $\eta$ is just through
$\bar{\eta}\eta$, it doesn't matter to put $\eta_1$ or $\eta_2$ in
the argument of $G$.

Now we will consider the surface-bulk Green function. Again,
because there are two $\eta$'s in the free action, this Green
function will relate $\Phi(x,\eta_2)$, which lives in the bulk, to
$\Phi_b(\vec{y},\eta_1)$, which lives on the boundary:
\begin{equation}\label{Kxy}
\Phi(x,\eta_{1})= \int d\bar{\eta}_{2}d\eta_{2}\int d^d \vec{y}
\,{\mathcal K}
(x,\vec{y},\eta_{1},\eta_{2})\Phi_{b}(\vec{y},\eta_{2}).
\end{equation}
Again, one can see that this Green function, is just the same as
the one derived in \cite{MRSAdS}, multiplied by
$\delta(\eta_1+\eta_2)$, that is:
\begin{equation}\label{formofK}
{\mathcal K}(x,\vec{y},\eta_{1},\eta_{2})=
K(x,\vec{y},\eta_{1})\delta(\eta_{1}+\eta_{2})
=a(\eta_{1})\left(\frac{x^{d}}{(x^{d})^{2} +|\vec{x}-\vec{y}|^{2}}
\right)^{\Delta+\bar{\eta}_{1}\eta_{1}}\delta(\eta_{1}+\eta_{2})
\end{equation}
with
\begin{equation}\label{aa'}
a(\eta_{1})=\frac{\Gamma(\Delta+\bar{\eta}_{1}\eta_{1})}
{2\pi^{d/2}\Gamma(\alpha+1)} =a+a^{\prime}\bar{\eta}_{1}\eta_{1},
\end{equation}
where $\alpha=\Delta+\bar{\eta}_{1}\eta_{1}-d/2$.

So far, we have derived Feynman rules for different propagators,
the last step to make the discussion complete is derivation of the
vertex term, which is present due to presence of interaction in
the theory. As the interaction term is proportional to product of
$n$ $\Phi$'s, to each vertex $n$ line's are attached, each of them
carrying a different $\eta$. The $\eta$'s are different because
they are different in the interaction term. The only condition
which should be satisfied, is that the sum of all $\eta$'s of the
vertex should vanish. This is done by putting proper delta
functions on the vertex. The strength of the interaction is
controlled by the coupling constant('s) $\lambda(\eta_i)$. Note
that this is not a simple coupling constant, one should expand it
in terms of different $\eta$'s (and $\bar{\eta}$'s) to have the
complete form.

\begin{figure}[t]
\begin{picture}(100,200)(0,0)
\includegraphics{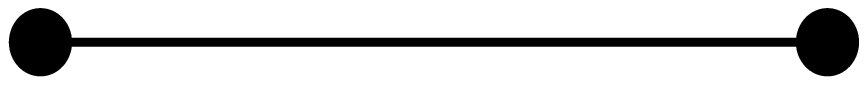} \put(40,193){$x$} \put(125,193){$y$}
\put(40,133){$\vec{x}$}\put(122,133){$y$} \put(60,203){$\eta_1$}
\put(106,203){$\eta_2$}
\put(58,141){$\eta_1$}\put(100,141){$\eta_2$}
\includegraphics{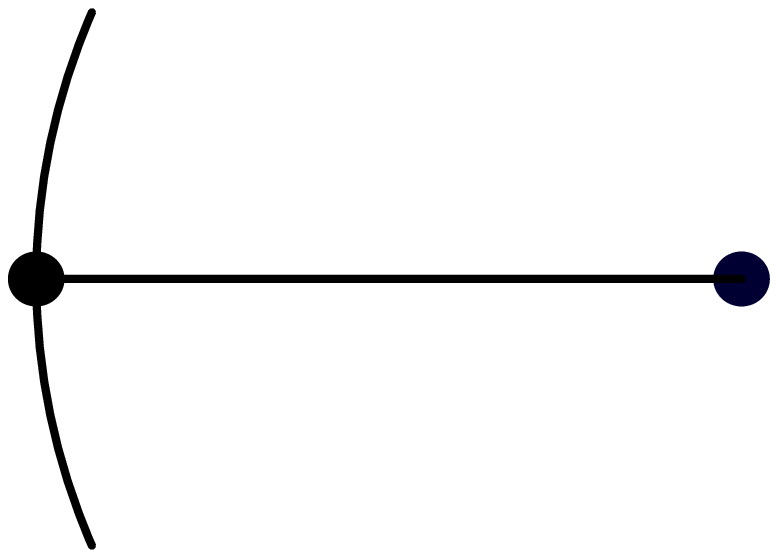} \includegraphics{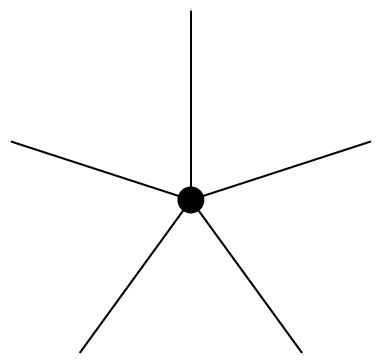} \put(72,92){$\eta_1$}\put(38,60){$\eta_2$}
\put(51,22){$\eta_3$}\put(108,22){$\eta_4$}\put(123,60){$\eta_5$}
\put(81,41){$x$}%\put(92,47){$\lambda(\eta_i)$}
\put(195,192){$\Rightarrow$}\put(195,132){$\Rightarrow$}
\put(195,50){$\Rightarrow$} \put(5,192){a)}\put(5,132){b)}
\put(5,50){c)} \put(245,192){$G(x,y,\eta_1)\delta(\eta_1+\eta_2)$}
\put(245,132){$\Phi_b(\vec{x},\eta_1)K(\vec{x},y,\eta_1)\delta(\eta_1+\eta_2)$}
\put(245,50){$\int d^{d+1}x \int
d{\bar\eta}_id\eta_i\sqrt{|g(x)|}\lambda(\eta_i)\delta(\sum\eta_i)$}
%\put(182,88){$x$} \put(182,119){$\lambda$}
\end{picture}
\caption{Different Feynman rules of the theory: a) Bulk-bulk
propagator,  b) Boundary-bulk propagator, c) Interaction vertex
(assuming the interaction term is proportional to $\Phi^5$)}
\end{figure}

We have briefly written the complete Feynman rules in figure 1. It
worths to mention that to every point, a spacial coordinate is
assigned, and to every ending of any propagator a grassmann
variable is assigned. Like all Feynman rules, one should integrate
over all undetermined spacial coordinates, in addition, one should
integrate over all undetermined grassmann variables.

To see how powerful this method is, we will calculate two point
function at tree level by considering following Feynman diagram
(figure 2). This diagram is present in every interactive theory,
in fact there is no interaction term present in the diagram. So
the result should turn out to be the same as derived in free
theory \cite{Khorrami,KoganAds,MRSAdS}. In \cite{MRSAdS} this was
done using grassmann variables, but the final result was not in
closed form.

\begin{figure}[t]
\begin{picture}(100,130)(0,0)
\includegraphics{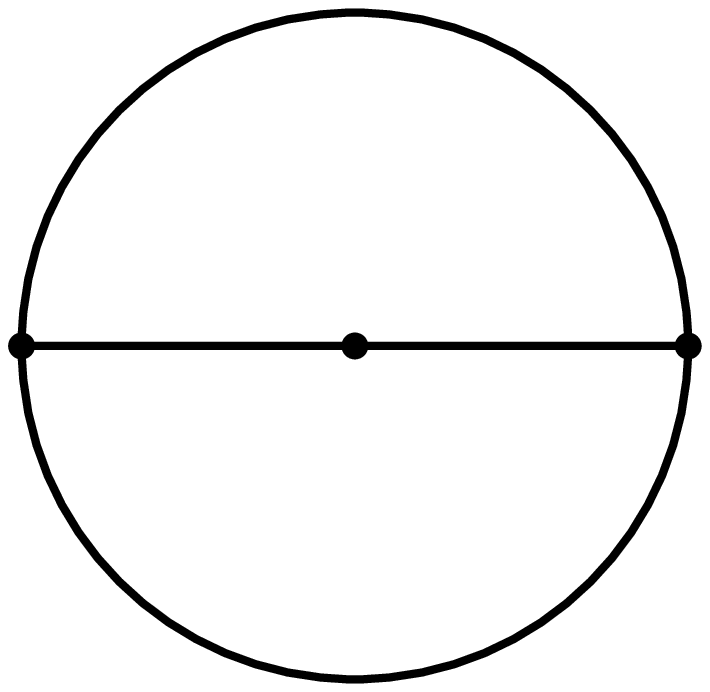}  \put(143,68){$\vec{y}_1$}
\put(200,58){$x$}\put(258,68){$\vec{y}_2$} \put(159,75){$\eta_1$}
\put(187,75){$\eta'_1$}\put(212,75){$\eta'_2$}\put(238,75){$\eta_2$}
\end{picture}
\caption{The tree level Feynman rule for two point function.}
\end{figure}

The term corresponding to the above Feynman diagram is read as
\begin{eqnarray}
\Phi_b(\vec{y}_1,\eta_1)\Phi_b(\vec{y}_2,\eta_2) \int d^{d+1} x
\sqrt{|g(x)|}\int d\bar{\eta}'_1d\eta'_1 d\bar{\eta}'_2d\eta'_2 \times\hspace{3cm}\nonumber\\
\hspace{1cm}K(\vec{y}_1,x,\eta_1)K(\vec{y}_2,x,\eta_2)
\delta(\eta_1+\eta'_1)\delta(\eta_2+\eta'_2)\delta(\eta'_1+\eta'_2).
\end{eqnarray}
Integrating over grassmannian variables leads to a single delta
function on external $\eta$'s, that is $\delta(\eta_1+\eta_2)$.
This is a general result and guarantees BRST symmetry. We will
discuss it more completely in the next section. There will remain
a single integral over spatial coordinate. By Ads/CFT
correspondence, the result should be equal to the two point
function in the CFT part multiplied by the value of boundary
fields: $\langle
\hat{O}(\vec{y}_1,\eta_1)\hat{O}(\vec{y}_2,\eta_2)\rangle
\Phi_b(\vec{y}_1,\eta_1)\Phi_b(\vec{y}_2,\eta_2)$.\footnote{In
fact, you have integrations over external coordinates and
grassmanns on both side of equation, but one can drop them without
loss of generality} The integration over $x$ can be done
\cite{MRSAdS} and the result is
\begin{equation}\label{2Point}
\langle \hat{O}(\vec{y}_1,\eta_1)\hat{O}(\vec{y}_2,\eta_2)\rangle=
\frac{a(\eta_1)}{|\vec{y}_1-\vec{y}_2|^{2\Delta+\bar{\eta}_1\eta_1}}\delta(\eta_1+\eta_2).
\end{equation}
The result is consistent with the one derived before
\cite{Khorrami,KoganAds,MRSAdS}, in addition, we have found the
correlation function in closed form. By expanding (\ref{2Point})
in powers of $\eta_i$'s, different correlation functions of the
fields $A$, $B$, $\bar{\zeta}$ and $\zeta$ are obtained.

\section{n-Point Correlation Functions at Tree Level}
When there is some kind of interaction in the bulk theory, the
n-point correlation function will find corrections. In this
section, we will find the correction to the first order of
$\lambda$ or equivalently to tree level. Let's consider an
$n$-point correlation function. In the framework of Witten
diagrams, this means that we should have $n$ distinct point on the
boundary. Each line should connect to others to produce a
connected diagram.  The simplest way to do that, is to take an
interaction action proportional to $\Phi^n$. In this case, one can
connect all the points on the boundary to an $n$-vertex in the
bulk (see figure 3 for the case n=4).

This diagram is equivalent to a term proportional to $\Phi_b^n$.
On the other hand,  by expanding exponential in the left hand of
equation (\ref{correspondence}), there will be a term having the
same factor in the following form:
\begin{eqnarray}\label{npoint}
&&\frac{1}{n!}\int d\vec{y}_{1}\cdots d\vec{y}_{n}\int
d\bar{\eta}_{1}^{\prime}d\eta_{1}^{\prime}\cdots
d\bar{\eta}_{n}^{\prime}d\eta_{n}^{\prime}\left\langle\hat{O}(\vec{y}_{1},\eta_{1}^{\prime})\cdots\hat{O}(\vec{y}_{n},\eta_{n}^{\prime})\right\rangle\nonumber\\
&&\hspace{6.2cm}\Phi_{b}(\vec{y}_{1},\eta_{1}^{\prime})\cdots\Phi_{b}(\vec{y}_{n},\eta_{n}^{\prime}).
\end{eqnarray}
\begin{figure}[h]
\begin{picture}(100,170)(-18,25)
\includegraphics{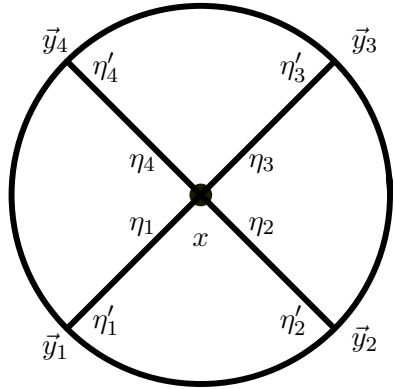} \put(182,88){$x$} %\put(182,119){$\lambda$}
\put(125,47){$\vec{y}_1$}\put(242,50){$\vec{y}_2$}\put(242,163){$\vec{y}_3$}\put(125,163){$\vec{y}_4$}
\put(144,57){$\eta'_1$}\put(215,57){$\eta'_2$}\put(215,153){$\eta'_3$}\put(144,153){$\eta'_4$}
\put(158,94){$\eta_1$}\put(203,94){$\eta_2$}\put(203,118){$\eta_3$}\put(158,118){$\eta_4$}
\end{picture}
\caption{The tree level correction to correlation function in the
case $n=4$}
\end{figure}
\vspace{2cm} So, using the Feynman rules, the correction to
$n$-point correlation function of the operators
$\hat{O}(\vec{y},\eta^{\prime})$ is read to be
\begin{eqnarray}\label{npointfunc}
\left\langle\hat{O}(\vec{y}_{1},\eta_{1}^{\prime})\cdots\hat{O}(\vec{y}_{n},\eta_{n}^{\prime})\right\rangle=
\int d^{d+1}x\sqrt{|g(x)|}\int d\bar{\eta}_{1}d\eta_{1}\cdots
d\bar{\eta}_{n}d\eta_{n}\delta(\eta_{1}+\cdots+\eta_{n})\nonumber\\
\lambda(\eta_{1},\cdots,\eta_{n})K(x,\vec{y}_{1},\eta_{1})\cdots
K(x,\vec{y}_{n},\eta_{n})\delta(\eta_1+\eta'_1)\cdots
\delta(\eta_n+\eta'_n)
\end{eqnarray}
The integrations on grassmann variables can be done easily, the
result is a single delta function on the external $\eta$'s. After
substituting $K(x,\vec{y}_{i},\eta_{i})$ from Eq. (\ref{formofK}),
we obtain n-point correlation function at tree level
\begin{eqnarray}
\left\langle\hat{O}(\vec{y}_{1},\eta_{1}^{\prime})\cdots\hat{O}(\vec{y}_{n},\eta_{n}^{\prime})\right\rangle
=a(\eta_{1}^{\prime}) \cdots a(\eta_{n}^{\prime})\int
d^{d+1}x\sqrt{|g(x)|}\delta(\eta_{1}^{\prime}+\cdots+\eta_{n}^{\prime})\nonumber\\
\lambda(\eta_{1}^{\prime},\cdots,\eta_{n}^{\prime})\left(\frac{x^{d}}{(x^{d})^{2}
+|\vec{x}-\vec{y}_{1}|^{2}}
\right)^{\Delta+\bar{\eta}_{1}^{\prime}\eta_{1}^{\prime}}\cdots\left(\frac{x^{d}}{(x^{d})^{2}
+|\vec{x}-\vec{y}_{n}|^{2}}
\right)^{\Delta+\bar{\eta}_{n}^{\prime}\eta_{n}^{\prime}}.
\end{eqnarray}
We would like to have some comment on this correction. First of
all, it is easy to check that the result is conformal covariant.
You should just note that the functions $K$ transform properly
under scaling, just you should note that it is possible to rescale
the dummy variable $x$. Second, note that there exists a delta
function on sum of external grassmann variables. As said before,
it guarantees BRST symmetry. Let's be more precise, consider the
n-point correlation function
$X=\left\langle\hat{O}(\vec{y}_{1},\eta_{1})
\cdots\hat{O}(\vec{y}_{n},\eta_{n})\right\rangle$. Under BRST
transformation,
$\Phi(\eta)\,\rightarrow\,(1+\bar{\epsilon}\eta)\Phi(\eta)$, we
have
\begin{equation}
\delta X= \bar{\epsilon}(\eta_1+\cdots+\eta_n) X.
\end{equation}
Due to BRST symmetry, the above expression should vanish. So, if
the sum of $\eta$'s is non zero, the correlator should vanish.
Consequently, the correlator should have a delta function on
$\eta$'s as a multiplier.

The other point to be clarified is that our coupling constant is a
compound one. We should expand it in terms of $\eta$'s (and
$\bar{\eta}$'s) to find the components. As an example we consider
the $\Phi^3$ theory. The coupling constant is written as
\begin{equation}\label{lambda}
\lambda(\eta_1,\eta_2,\eta_3,\bar{\eta}_1,\bar{\eta}_2,\bar{\eta}_3)=
\lambda_0 + \bar{\eta}_i\eta_j \lambda_{ij}+
\bar{\eta}_i\eta_j\bar{\eta}_k\eta_l \lambda_{ijkl}+
\bar{\eta}_i\eta_j\bar{\eta}_k\eta_l \bar{\eta}_m\eta_n
\lambda_{ijklmn}.
\end{equation}
There is no other term in this series because our variables are
grassmanns. Further investigation shows that the last term is
irrelevant. This happens because we have a delta function on
$\eta'$s which imposes a constraint on the garassmans. So one of
the grassmann can be written in terms of the others and the last
term in equation (\ref{lambda}) vanishes. Also in the expansion of
$\lambda$ we have not written terms like $\bar{\beta}_i\eta_i$ as
there should not be any grassmannian constant in the theory.

With this coupling constant('s) we can find out what vertices are
present. As the last term is absent, we do not have the $CCC$
vertex. This leads to vanishing correlator of $\langle
\hat{A}\hat{A}\hat{A} \rangle$ in CFT part, which is a desired
result. Also vertices which have odd number of grassmann fields do
not exist, but other vertices are present in the theory.

Let's investigate the $\Phi^3$ theory more precisely and derive
the three point functions explicitly. This is a good check for our
theory, as we know the form of correlators from symmetry. Also,
knowing the three point functions, one can say something about
OPE's of the fields present in the theory.

In this case the integrations can be done explicitly
\begin{eqnarray}\label{3point}
&&\langle
\hat{O}(\vec{y}_{1},\eta_{1})\hat{O}(\vec{y}_{2},\eta_{2})\hat{O}(\vec{y}_{3},\eta_{3})\rangle=\nonumber\\
&&a(\eta_{1})a(\eta_{2})
a(\eta_{3})\delta(\eta_{1}+\eta_{2}+\eta_{3})\lambda(\eta_{1},\eta_{2},\eta_{3})
\frac{\pi^{\frac{d}{2}}}{2}\frac{\Gamma\left(\frac{3\Delta}{2}-\frac{d}{2}+\frac{\bar{\eta}_{1}\eta_{1}+
\bar{\eta}_{2}\eta_{2}+\bar{\eta}_{3}\eta_{3}}{2}\right)}{\Gamma(\Delta+\bar{\eta}_{1}\eta_{1})\Gamma(\Delta+\bar{\eta}_{2}\eta_{2})
\Gamma(\Delta+\bar{\eta}_{3}\eta_{3})}\nonumber\\
&&\frac{\Gamma\left(\frac{\Delta}{2}+\frac{-\bar{\eta}_{1}\eta_{1}+
\bar{\eta}_{2}\eta_{2}+\bar{\eta}_{3}\eta_{3}}{2}\right)
\Gamma\left(\frac{\Delta}{2}+\frac{\bar{\eta}_{1}\eta_{1}-
\bar{\eta}_{2}\eta_{2}+\bar{\eta}_{3}\eta_{3}}{2}\right)\Gamma\left(\frac{\Delta}{2}+\frac{\bar{\eta}_{1}\eta_{1}+
\bar{\eta}_{2}\eta_{2}-\bar{\eta}_{3}\eta_{3}}{2}\right)}{y_{12}^{\Delta+\bar{\eta}_{1}\eta_{1}+
\bar{\eta}_{2}\eta_{2}-\bar{\eta}_{3}\eta_{3}}
y_{13}^{\Delta+\bar{\eta}_{1}\eta_{1}-
\bar{\eta}_{2}\eta_{2}+\bar{\eta}_{3}\eta_{3}}y_{23}^{\Delta-\bar{\eta}_{1}\eta_{1}+
\bar{\eta}_{2}\eta_{2}+\bar{\eta}_{3}\eta_{3}}},
\end{eqnarray}
where $y_{ij}=|\vec{y}_i-\vec{y}_j|$. Expanding both sides of the
above equation, different correlators are obtained. The
expressions are very huge and we will bring here only some of them
\begin{eqnarray}
&&\langle
\hat{A}(\vec{y}_{1})\hat{A}(\vec{y}_{2})\hat{B}(\vec{y}_{3})\rangle=-\langle
\hat{A}(\vec{y}_{1})\hat{\bar{\zeta}}(\vec{y}_{2})\hat{\zeta}(\vec{y}_{3})\rangle
=\lambda_{0}b_1\frac{1}{y_{12}^{\Delta}y_{13}^{\Delta}y_{23}^{\Delta}}\nonumber\\
&&\langle
\hat{\zeta}(\vec{y}_{1})\hat{\bar{\zeta}}(\vec{y}_{2})\hat{B}(\vec{y}_{3})\rangle
=\left(\frac{a^{\prime}}{a}\lambda_{0}+\lambda_{12}
+\lambda_{13}+\lambda_{32}+\lambda_{33}\right)b_1
\frac{1}{y_{12}^{\Delta}y_{13}^{\Delta}y_{23}^{\Delta}}\nonumber\\
&&\hspace{3.4cm}+\lambda_{0}
\frac{1}{y_{12}^{\Delta}y_{13}^{\Delta}y_{23}^{\Delta}}\left(b_2+b_3
\ln\frac{y_{13}y_{23}}{y_{12}}\right)
\end{eqnarray}
where $b_1$, $b_2$ and $b_3$ are some constants depending on
$\Delta$, and $a$ and $a'$ are defined in equation (\ref{aa'}).
Other correlation functions have similar structure and are
consistent with previous results \cite{MRS,MRSAlgeb}.
%Since $\eta$ and $\bar{\eta}$ are grassmann variables we can write
%$\delta(\eta'_{1}+\cdots+\eta'_{n})=\eta'_{1}+\cdots+\eta'_{n}$,
%then
\section{Loops}
In this last section, we briefly investigate the general
properties of arbitrary generalized Weitten diagrams which may
contain loops. Though nearly none of them can be calculated
explicitly, still there is a lot to say. As an example consider
the $\Phi^4$ and the one loop correction to four point function.
The corresponding diagram is shown in figure 4.
\begin{figure}[t]
\begin{picture}(100,170)(0,0)
\includegraphics{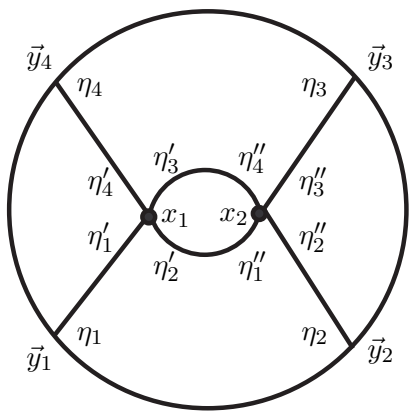} \put(176,81.5){$x_1$}\put(198,82){$x_2$} %\put(182,119){$\lambda$}
\put(125,27){$\vec{y}_1$}\put(254,30){$\vec{y}_2$}\put(254,141){$\vec{y}_3$}\put(125,141){$\vec{y}_4$}
\put(144,37){$\eta_1$}\put(229,37){$\eta_2$}\put(229,131){$\eta_3$}\put(144,131){$\eta_4$}
\put(148,73){$\eta'_1$}\put(173,62){$\eta'_2$}\put(173,103){$\eta'_3$}\put(148,94){$\eta'_4$}
\put(205,62){$\eta''_1$}\put(228,73){$\eta''_2$}\put(228,94){$\eta''_3$}\put(205,103){$\eta''_4$}
\end{picture}
\caption{The one-loop correction to four point function in the
$\Phi^4$ theory.}
\end{figure}
The term associated to this diagram can be read easily: we have
four lines joining bulk to boundary, two lines in the bulk and two
vertices. The expression is too lengthy, so we will not bring it
here.

Most of  integrations on $\eta'_i$s and $\eta''_j$s can be done
easily due to presence of delta functions. Just like before, the
over all result has a delta function on sum of external grassmann
variables, which means BRST invariance. This leads to vanishing
correlators which have only $\hat{A}(\vec{y})$. To see how this
happens, it is sufficient to note that the delta functions on
grassmann variables can be written in the following form
\begin{equation}
\delta(\eta_1+\eta_2)=\eta_1+\eta_2.
\end{equation}
So, in front of the correlator, there stands a factor like $\sum
\eta_i$ and there is no term in the correlator that has no
dependence on $\eta_i$. This is just the term consist of
$\hat{A}$'s, only. This result has been seen before
\cite{Flohr1,MRS}, and has origin in the fact that the expectation
value of the unity operator vanishes.

There will remain three integrations, two over spatial coordinates
$x_1$ and $x_2$, and the other over one grassmann variable (and
its complex conjugate) which play a role like the circulating
momenta in the usual Feynman diagrams.

To check conformal invariance, one should know how the functions
$K$ and $G$ change under these transformations. Transformation of
$K$ is easily derived from it's form (equation (\ref{formofK})):
it rescales and its weight is $\Delta+\bar{\eta}\eta$. To see this
in the case of simple scaling, it is enough to rescale the dummy
variables $x_1$ and $x_2$. Under special conformal transformation,
with similar but more complicated calculation, one can show the
same result is obtained \cite{Laz}. The bulk-bulk propagators,
$G$, happen to be invariant under scalings. This is shown by
\cite{Laz} in ordinary CFT, generalization to LCFT is
straightforward, using grassmann variables. This shows that the
diagram sketched above, has conformal invariance. It is clear that
this happens to all diagrams of any order and any number of loops
and conformal symmetry is preserved in this loop expansion.

\end{document}